\documentclass[aps,prl,twocolumn,superscriptaddress]{revtex4}
\usepackage{amsmath}
\usepackage{amsfonts}
\usepackage{amssymb}
\usepackage{graphicx}
\usepackage{dcolumn}
\usepackage{bm}

\begin{document}

\title{Itinerant spin excitations near the hidden order transition in
URu$_{2}$Si$_{2}$}
\author{J.~A.~Janik}\email{janik@magnet.fsu.edu}
\thanks{\emph{Present address: Geophysical Laboratory, Carnegie Institution of
Washington, Washington, DC 20015}} \affiliation{Department of
Physics, Florida State University, Tallahassee, FL 32306-3016,
USA} \affiliation{National High Magnetic Field Laboratory, Florida
State University, Tallahassee, FL 32306-4005, USA}
\author{H.~D. Zhou}
\affiliation{National High Magnetic Field Laboratory, Florida
State University, Tallahassee, FL 32306-4005, USA}
\author{Y.-J.~Jo}
\affiliation{National High Magnetic Field Laboratory, Florida
State University, Tallahassee, FL 32306-4005, USA}
\author{L.~Balicas}
\affiliation{National High Magnetic Field Laboratory, Florida
State University, Tallahassee, FL 32306-4005, USA}
\author{G.~J.~MacDougall}
\affiliation{Department of Physics and Astronomy, McMaster
University, Hamilton, Ontario L8S 4M1, Canada}
\author{G.~M.~Luke}
\affiliation{Department of Physics and Astronomy, McMaster
University, Hamilton, Ontario L8S 4M1, Canada}
\affiliation{Canadian Institute for Advanced Research, Toronto,
Ontario, Canada M5G 1Z8}
\author{J.~D.~Garrett}
\affiliation{Brockhouse Institute for Materials Research, McMaster
University, Hamilton, Ontario L8S 4M1, Canada}
\author{K.~J.~McClellan}
\affiliation{Science, Technology \& Engineering Directorate, Los
Alamos National Laboratory, Los Alamos, New Mexico 87545, USA}
\author{E.~D.~Bauer}
\affiliation{Science, Technology \& Engineering Directorate, Los
Alamos National Laboratory, Los Alamos, New Mexico 87545, USA}
\author{J.~L.~Sarrao}
\affiliation{Science, Technology \& Engineering Directorate, Los
Alamos National Laboratory, Los Alamos, New Mexico 87545, USA}
\author{Y.~Qiu}
\affiliation{NIST Center for Neutron Research, Gaithersburg,
Maryland, 20899-6102, USA} \affiliation{Department of Materials
Science and Engineering, University of Maryland, College Park,
Maryland, 20742, USA}
\author{J.~R.~D.~Copley}
\affiliation{NIST Center for Neutron Research, Gaithersburg,
Maryland, 20899-6102, USA}
\author{Z. Yamani}
\affiliation{CNBC, National Research Council, Chalk River, ON K0J
1J0, Canada}
\author{W.~J.~L.~Buyers}
\affiliation{CNBC, National Research Council, Chalk River, ON K0J
1J0, Canada} \affiliation{Canadian Institute for Advanced
Research, Toronto, Ontario, Canada M5G 1Z8}
\author{C. R. Wiebe}
\affiliation{Department of Physics, Florida State University,
Tallahassee, FL 32306-3016, USA} \affiliation{National High
Magnetic Field Laboratory, Florida State University, Tallahassee,
FL 32306-4005, USA}
\date{\today }
\begin{abstract}
By means of neutron scattering we show that the high-temperature
precursor to the hidden order state of the heavy fermion
superconductor URu$_{2}$Si$_{2}$ exhibits heavily damped
incommensurate paramagnons whose strong energy dispersion is very
similar to that of the long-lived longitudinal f-spin excitations
that appear below T$_{0}$.  Since the underlying local f-exchange
is preserved we expect only the f-d interactions to change across
the phase transition and to cause the paramagnetic damping.  The
damping exhibits single-ion behavior independent of wave vector
and vanishes below the hidden order transition.  We suggest that
this arises from a transition from valence fluctuations to a
hybridized f-d state below T$_{0}$.  Here we present evidence that
the itinerant excitations, like those in chromium, are due to
Fermi surface nesting of hole and electron pockets so that the
hidden order phase likely originates from a Fermi-surface
instability. We identify wave vectors that span nested regions of
a band calculation and that match the neutron spin crossover from
incommensurate to commensurate on approach to the hidden order
phase.
\end{abstract}
 \maketitle

The heavy fermion superconductor URu$_{2}$Si$_{2}$ has puzzled
physicists for more than two decades \cite{Fisk, Chandra,
Maple,Broholm1,Broholm2} yet, despite recent progress
\cite{Wiebe2007}, theoretical and experimental work is still
required to unravel its mysteries. The large change in entropy
across the lambda-like anomaly at $T_{0}=17.5$K in
URu$_{2}$Si$_{2}$ denotes a phase transition has taken place, but
unlike other second order transitions, we can find no large order
parameter to accompany that change.  Neutron scattering
measurements of the staggered magnetic moment
\cite{Broholm1,Broholm2,Metoki2000}, show it is too small to
explain the specific heat jump and hidden order, but
incommensurate spin density waves have been detected that might
explain the entropy loss \cite{Wiebe2007}. Experiment finds that
spin excitations are strong and longitudinal and belong to the
bulk hidden order phase \cite{Broholm2}. Theoretical perspectives
on URu$_{2}$Si$_{2}$ generally fall into either localized or
itinerant scenarios.  In the first, f-electrons on the uranium
sites must have a quadrupolar or even octupolar character
\cite{Santini1994,Ohkawa1999}.  This scenario is not supported by
the absence of observable crystal field excitations
\cite{Broholm1,Broholm2,Wiebe2007}. In the second, a Fermi surface
instability leads to a restructuring and the subsequent change in
entropy \cite{Maple, Jeffries2007, Varma}.

Our recent work on URu$_{2}$Si$_{2}$ \cite{Wiebe2007} revealed
that spin fluctuations emanate from incommensurate wave vectors
with a relatively large velocity, that they can account for the
large specific heat, and can explain its reduction below T$_{0}$
by the gapping of the spin excitations at all wave vectors.  In an
earlier search for orbital currents, as rings of magnetic
incommensurate scattering in reciprocal space with a Q$^{-4}$ form
factor \cite{Chandra}, Wiebe \textit{et al.} \cite{Wiebe2004}
found that any such ring scattering was undetectable.  Instead,
they found enhanced thermally activated scattering at the
(0.6,0,0) position. From our high resolution NIST work
\cite{Wiebe2007} on the Disk Chopper Spectrometer (DCS) instrument
above and below the T$_{0}$ transition, we found dynamic
incommensurate scattering around the forbidden (1$\pm\delta$,0,0)
position with $\delta=0.4$ at $T=20$ K and concluded the
following: (1) these highly correlated and gapless excitations
have a well defined Q-structure and their incommensurability with
the lattice suggests itinerant electronic behavior as opposed to
localized electron physics, (2) a restructuring of the Fermi
surface must be responsible for the hidden order transition, and
(3) the gapping of the incommensurate excitations below T$_{0}$ in
the hidden order state removes thermally accessible excitations
and accounts for the missing entropy. The entropy was calculated
by fitting the over-damped modes to a spin density wave model
\cite{Chou,Stock}, extracting a correlation length, and using a
linear DOS counting argument. Despite not knowing precisely what
the hidden ordered state is, we believe that ref. \cite{Wiebe2007}
identified the collective modes associated with the hidden order.
Broholm \textit{et al.} \cite{Broholm2} used a model of two
singlets coupled by seven exchange constants to explain the
excitation spectrum.  We will demonstrate in this Letter how
nesting of the Fermi surface leads to the incommensurate wave
vector of the over-damped collective modes. Our findings
dramatically simplify the understanding of the excitations in the
system.

High purity U, Ru, and Si were melted in a mono-arc furnace into
URu$_{2}$Si$_{2}$ buttons. Three large single crystals, 20g in
total, were then grown via a modified Czochralski method in a
titanium gettered argon atmosphere tri-arc furnace followed by
annealing in argon at 900$^{o}$C. Three crystals were co-aligned
in the (H0L) scattering plane, two grown at Los Alamos National
Laboratory and one at McMaster University.  Each sample was
confirmed to be a single crystal in a $2\pi$ survey at the E3
spectrometer at Chalk River. We performed extensive inelastic
neutron scattering (INS) studies on URu$_{2}$Si$_{2}$ near the
hidden order transition temperature. In the first experiment the
setup was the same as in reference \cite{Wiebe2007}. In the second
experiment with the C5 Triple Axis Spectrometer at the NRU reactor
in Chalk River, Canada, we followed the excitations to larger
energies.  The instrument setup and collimation were set to
0.5$^{o}$-PG-0.80$^{o}$-S-0.85$^{o}$-PG-2.4$^{o}$ with final
scattering energy E$_{f}$=14.6 meV. Two graphite filters in the
scattered beam eliminated higher-order feed through. Constant-Q
scans were performed up to ~20 meV at wave vector transfers Q =
(1+h,0,0) for h = 0 to 1 and compared to measurements at (3-h,0,0)
to detect the presence of phonons. For $H<1.65$ the scattering is
entirely magnetic.  Constant energy scans up to 16.5 meV were
performed to determine the momentum width of the cones of
incommensurate scattering.
\begin{figure}[h!]
\begin{center}
\includegraphics[scale=0.5,angle=0]{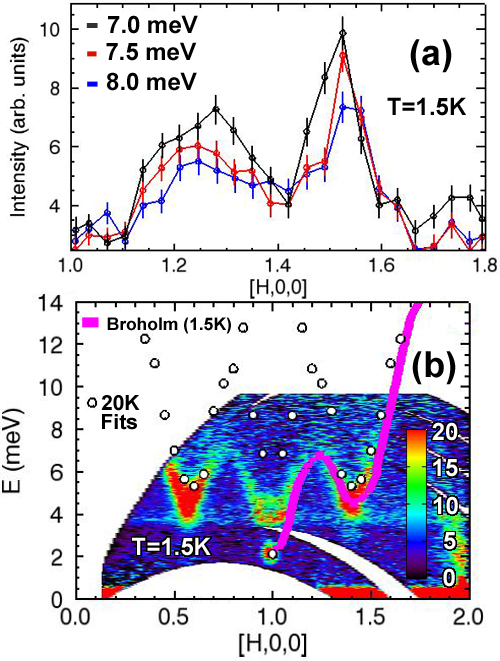}
\end{center}
\caption{High resolution DCS data (b) integrated in the L
direction from [-0.15 to 0.15], at $T=1.5$ K deep in the hidden
ordered state. We can see the excitations at the commensurate
(100) position and the modes previously thought of as a ``mode
softening" centered on the incommensurate positions
(1$\pm0.4$,0,0). Overlayed in magenta is the dispersion relation
from the model of Broholm \emph{et al.} \cite{Broholm2} from their
triple axis work. In panel (a), one can see the excitations extend
to higher energy than previously suggested.  The cuts are energy
integrated from [6.5,7.5], [7,8], [7.5,8.5] meV (with the same L
integration as (b)) in arbitrary units for the black, red, and
blue, traces respectively.  In (b) we overlay the 1.5 K data the
resonant paramagnon energy at 20 K from fits to the damped simple
harmonic oscillator spectrum to the thermal Chalk River data.}
\label{fig1}
\end{figure}

In Fig.~\ref{fig1} (b), we reproduce our high resolution NIST data
at $T=1.5K$, well into the hidden order state (the Q-independent
spurion at 3.9 meV is due to fast fission processes in uranium).
We can see the excitations above the commensurate (100) position
and the modes previously thought of as a ``mode softening"
centered on the incommensurate position (1$\pm0.4$,0,0). Overlayed
in Fig.~ \ref{fig1} (b) in magenta is the dispersion measured by
Broholm \emph{et al.} earlier \cite{Broholm2}.  The subtraction
routine used to remove the spurion \cite{Wiebe2007} also removes
intensity above H=1.3 which is clearly visible Fig.~\ref{fig1}
(b). One can see in Fig.~\ref{fig1} (a) 1 meV wide constant-energy
cuts along [H00]. These cuts near the local maxima of the Broholm
data show increased scattering above the single peak of their
model. Thus the excitations extend higher in energy than
previously thought. The broad tails to higher energy suggest that
the sharp spin peak is the onset of a continuum, rather than a
long-lived spin wave.
\begin{figure}[h!]
\begin{center}
\includegraphics[scale=0.5,angle=0]{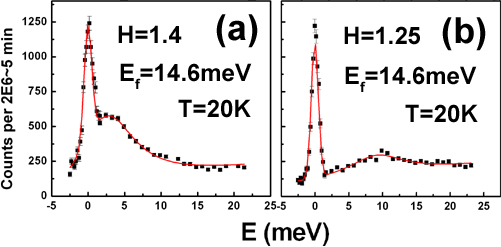}
\end{center}
\caption{The constant-Q scans at $T=20$ K for $\textbf{Q}=(H,0,0)$
in (a) and (b) show how the energy of the damped paramagnon
resonance disperses with wave vectors.  The lines are a damped
simple harmonic oscillator fit whose resonant energy is plotted in
Fig. \ref{fig1} (b).  Note the presence of damped ungapped spin
fluctuations existing up to high energies.} \label{fig2}
\end{figure}

In Fig.~\ref{fig2} (a) and (b) we present examples of constant-Q
scans with thermal neutrons at NRU above the hidden order phase at
$T=20$ K.  The paramagnetic scattering, although broad in energy,
shows a peak whose energy rises from minima at H=0 and H=1.4, the
wavevectors whose magnetic susceptibility is largest.  We fit the
inelastic spectrum to a damped simple harmonic oscillator model
and to an elastic gaussian plus a sloping background. The
inelastic spectrum is given by
\begin{equation}
S(\omega ,Q)=(n\left( \omega \right) +1)\frac{A\omega \omega _{c}\Gamma }{%
\left( \omega ^{2}-\omega _{c}^{2}\right) ^{2}+\omega ^{2}\Gamma
^{2}}.
\end{equation}
where $n\left( \omega \right) =
(\exp{(\frac{\hbar\omega}{k_{B}T})}-1)^{-1}$ is the Bose factor,
$\Gamma$ is the damping, and $\omega _{c}$ is the resonant energy.
We found that a Q-independent damping rate of  $\Gamma=10\pm1$ meV
best describes the data.  In Fig.~\ref{fig1} (b) the resonant
energy of the paramagnetic scattering is plotted as full circles
and is folded about (100).  The paramagnon dispersion at 20 K is
found to be surprisingly similar to that of the long-lived
longitudinal spin excitations of the hidden order phase at 1.4 K.
The main difference is that the paramagnons are heavily damped.
However, there remain energy minima at 1.0 and the incommensurate
1.4 position that generate two maxima in the susceptibility
$A/\omega_{c}(Q)$. The existence of a precursor phase of strongly
damped excitations is reminiscent of other itinerant electron
systems near magnetic ordering transitions such as chromium and
MnSi and can be represented as a general result of self-consistent
renormalization theory \cite{Moriya}. In addition to the broad
spectrum of damped paramagnons we also found that a commensurate
central mode appeared for $\textbf{Q}\rightarrow(1,0,0)$ in scans
such as Fig.~\ref{fig2} consistent with the commensurate precursor
at 20 K \cite{Broholm2}. This indicates that slow fluctuations of
the hidden order phase take place on approach to $T_{0}$.

\begin{figure}[h!]
\begin{center}
\includegraphics[scale=0.3,angle=0]{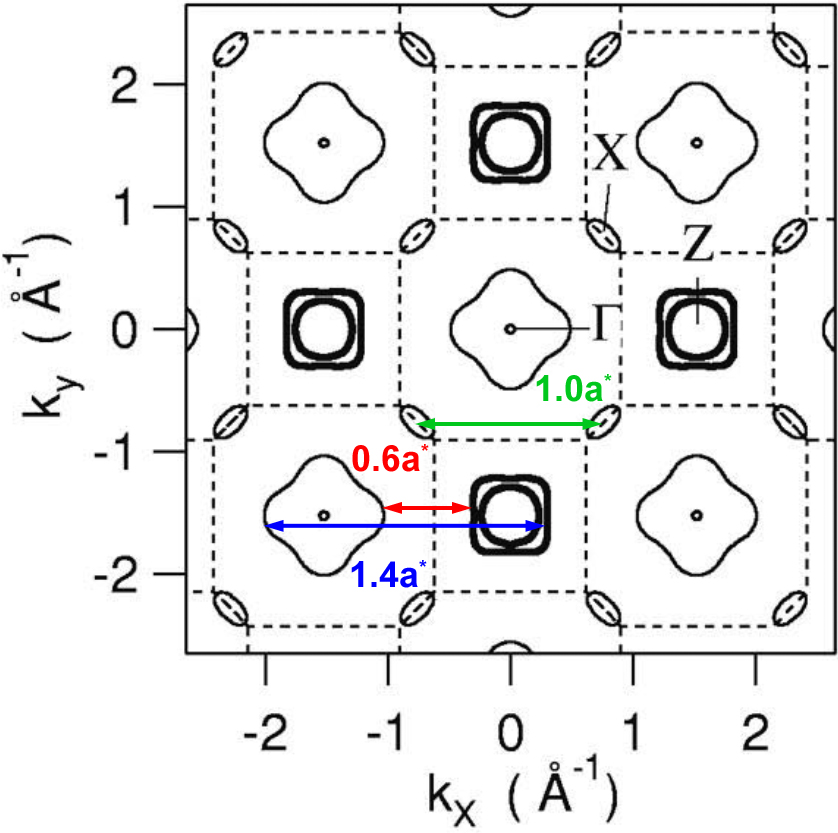}
\end{center}
\caption{The calculated Fermi surface of URu$_{2}$Si$_{2}$
\cite{Denlinger} and possible nesting vectors which may produce
the commensurate and incommensurate itinerant excitations.}
\label{fig3}
\end{figure}

We now present a new interpretation of the spin wave spectrum
above and below T$_{0}$ in terms of nesting at the Fermi surface.
The Fermi surface of URu$_2$Si$_2$ from recent band structure
calculations \cite{Denlinger} is reproduced in Fig.~\ref{fig3}. We
see that the incommensurate wavevectors of 0.6 a* and 1.4 a* can
come from the nesting of the electron jack at $\Gamma$ with the
two hole pockets at Z. We can eliminate several other
possibilities based upon the topology of the Fermi surface (which
must have nearly parallel sheets for nesting), and also with the
knowledge that the excitations are along the a* and b* directions.
While wave vector matching does not prove nesting (the velocity at
Z is larger than at $\Gamma$) it is at least suggestive that the
incommensurate excitation mode is represented by an electron
``jack" - hole pocket excitation above T$_{0}$. Below T$_{0}$, the
electron ellipse nesting vectors X-X$^{'}$ may lead to the
commensurate excitation at the (100) position.  The commensurate X
to X$^{'}$ magnetic transition involves an X-point ARPES density
that grows rapidly on cooling (see \cite{Denlinger} Fig 18.) and
might explain the crossover from incommensurate to commensurate
susceptibility \cite{Buyers1994} on cooling below 23 K. The ARPES
data \cite{Denlinger} was measured at high temperatures (T = 200
K). To confirm our nesting hypothesis would require accurate
measurements at temperatures of order 20 K.  There is not a
perfect nesting condition for the electron pockets, but this may
be an artifact of the calculation, and the pockets could actually
be more symmetric in shape. However, the matching of these
wavevectors is quite good and given the arguments for itinerant
magnetism stated above, it is reasonable to assume that nesting
plays an important role in this transition. What is unclear is why
these electron pocket nesting vectors (associated with the Q =
(1,0,0) wavevector) dominate in the hidden ordered state for $T <
T_{0}$.  We speculate that on cooling the f weight may move to a
lower hybrid band with commensurate nesting, whereas above the
hidden order phase the valence fluctuations of the thermally
broadened upper hybrid bands are nested incommensurately.  A
recent experimental paper suggests that the existence of
commensurate (1,0,0) excitations is a requirement for the
formation of Cooper pairs at low temperatures \cite{Villaume}.
\begin{figure}[h!]
\begin{center}
\includegraphics[scale=0.6,angle=0]{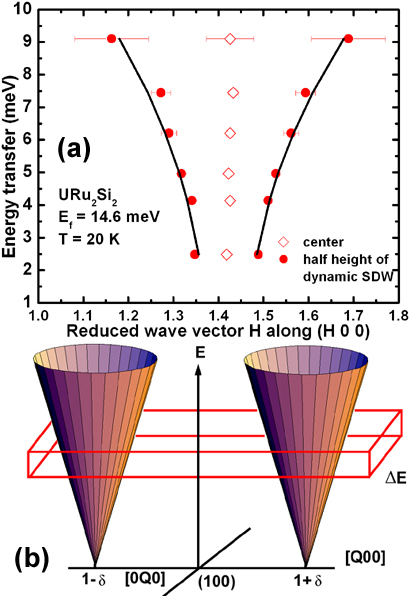}
\end{center}
\caption{In the upper panel (a) from constant-energy scans we show
the spreading with energy of the width in wave vector of the
incommensurate dynamic spin density wave at $T=20$ K. The lower
panel (b) shows a cartoon of the incommensurate fluctuations in
the H-K plane.} \label{fig4}
\end{figure}

Our data for URu$_{2}$Si$_{2}$ shows some similarities to the
itinerant system chromium in that a SDW forms at a nesting wave
vector \cite{Cr}. In URu$_{2}$Si$_{2}$, however, the
incommensurate fluctuations are dynamic as opposed to the static
SDW of chromium. Only under pressure does a static spin order of
0.3 $\mu_{B}$ occur, and then only at the commensurate (100)
wavevector \cite{Bourdarot}.  In Fig.~\ref{fig4} (a) we show
growth with energy of the the FWHH in wave vector of the
incommensurate filled cones of scattering at T$=20$K and (b) a
cartoon depicting how the filled cones evolve with energy. The
cones of scattering observed in URu$_2$Si$_2$ are very similar to
the fast itinerant excitations in Cr apart from the fact that the
incommensurate scattering in URu$_2$Si$_{2}$ forms a solid cone
\cite{Wiebe2007}.

While a full explanation of the URu$_{2}$Si$_{2}$ hidden order
state still remains to be found, we have laid out a consistent
explanation for how the excitation spectrum relates to the Fermi
surface of itinerant electrons.  We find that the paramagnon
dispersion above T$_{0}$ follows the spin wave dispersion for T
$\leq$ T$_{0}$, so it follows that the exchange interaction
between uranium moments does not change across T$_{0}$.  Varma and
Zhu claim \cite{Varma} that with helical or Pomeranchuk order,
there will be a gap opening with a reduction in the low energy
transverse spectral weight. Although they are non-specific about
the nature of the change in spectrum, our experiment shows that
the response arises only from longitudinal fluctuations. The main
characteristic of the phase transition is that the electronic
damping of longitudinal fluctuations decreases dramatically across
T$_{0}$.

This work was supported by the NSF CDMR-0084173, DMR-0454672, the
EIEG program (FSU) and the State of Florida.  The work at McMaster
is supported by NSERC  and at Los Alamos by the US DOE.  The
authors are grateful for the local support staff at the NIST
Center for Neutron Research and Chalk River Laboratories.

\end{document}